\documentclass[12pt]{article}
\usepackage{axodraw}
\usepackage{graphicx}
\begin{document}
\newcommand{\bq}{\begin{equation}}
\newcommand{\eq}{\end{equation}}
\newcommand{\bqa}{\begin{eqnarray}}
\newcommand{\eqa}{\end{eqnarray}}
\newcommand{\nll}{\nonumber\\}
\newcommand{\Litwo}{\mbox{${\rm{Li}}_{2}$}}
\newcommand{\als}{\alpha_{_S}}
\newcommand{\alsS}{\alpha^2_{_S}}
\newcommand{\sss}[1]{\scriptscriptstyle{#1}}
\newcommand{\ds }{\displaystyle}
\def\mw {M_{\sss{W}}}
\def\mtp{m_t}
\def\mbt{m_b}
\def\mg{m_g}
\def\mwt {\widetilde{M}^2_{\sss{W}}}
\def\thle{\vartheta_{l}}

%  \begin{flushright}
%{{\tt hep-ph/yymmnnn}\\
%      October 2006 
%}
%\end{flushright}
%\vspace*{3cm}
\begin{center}

{\Large\bf QCD branch in {\tt SANC}}
\vspace*{.1cm}

{\Large\bf (Implementation of NLO QCD corrections into
            framework of computer system {\tt SANC})}
\vspace*{.6cm}

{\bf A. ~Andonov$^1$, A.~Arbuzov$^{2,3}$, S.~Bondarenko$^{2,3}$, \\[1mm]
P.~Christova$^3$, V.~Kolesnikov$^3$, R.~Sadykov$^3$ }
\vspace*{8mm}

{\normalsize
{\it 
$^{1}$Bishop Konstantin Preslavsky University, Shoumen, Bulgaria\\ 
$^{2}$Bogoliubov Laboratory of  Theoretical Physics, JINR \\ 
$^{3}$Dzhelepov Laboratory for Nuclear Problems, JINR     \\
       ul. Joliot-Curie 6, RU-141980 Dubna, Russia }}
\vspace*{8mm}

\end{center}

\begin{abstract}
\noindent
The QCD sector of the system {\tt SANC} is presented.
QCD theoretical predictions for several processes of high energy interactions 
of fundamental particles at the one-loop precision level for up to 
some 3- and 4-particle processes are implemented.

\end{abstract}

\section{Introduction}

    The computer system {\tt SANC} is aimed to carry out semi--automatic 
calculations at the one-loop precision level of realistic and pseudo--observables 
for various processes of elementary particle interactions to be investigated  
at the present and future colliders -- Tevatron, LHC, ILC and others.

    We created the QCD environment of  {\tt SANC} and started
to implement systematically NLO QCD processes  
filling the {\bf QCD} branch of {\tt SANC} tree
in the same spirit as in Ref.~\cite{SANC1.00}.

   We created a set of FORM~\cite{Vermaseren} procedures for the analytic 
calculation of  building blocks of  QCD such as self-energies of 
quarks and gluons, vertices with virtual gluons and corresponding 
counter terms. These blocks are placed into the {\bf QCD} Precomputation level
of the system. The FORM programs are  accessible via the  same menu sequences 
as for {\bf QED} or {\bf EW} Precomputation.

\begin{figure}[!ht]
\includegraphics{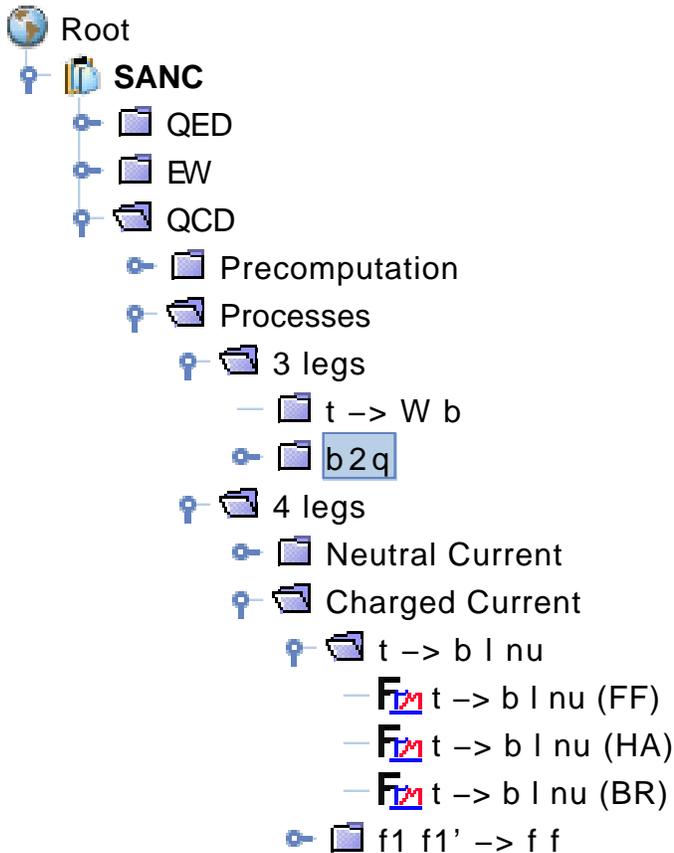}
\caption {QCD part of the {\tt SANC} tree.}
\label{QCDtree}
\end{figure}

   We consider here the results of implementation of the first processes  
of the {\bf QCD} branch  available in {\tt SANC}.
They are subdivided into {\bf 3legs} and {\bf 4legs} branches as one can see in  
Fig.~\ref{QCDtree}.
The  {\bf 3legs} branch contains  {\bf b2q} decays, namely: $t \to W\, b$, 
 $W \to u \, {\bar d}$, $W \to c \, {\bar s}$, $Z \to q \, {\bar q} $ and 
$H \to q \, {\bar q} $.
Here {\bf b} and {\bf q} denote any weak {\tt boson} and any {\tt quark},
 respectively, b denotes the b-quark.

The {\bf 4legs} branch contains  {\bf 4f} processes.
For the latter  there is a branch for {\bf Neutral Current} (NC) processes that
contains  the Drell--Yan process
$q \,{\bar q} \to \ell^- \, \ell^+ $, and 
a branch for {\bf Charged Current} (CC) processes that contains 
the decay $t\to b\,\ell^+\,\nu_\ell$ and the Drell--Yan $2f\to 2f$ process
 ${u}\, \bar{d}\to \ell^+\,{\nu}_\ell$.

 The structure of these branches 
is the same as the corresponding structure  in
the {\bf EW} sector of {\tt SANC}. For each process there are three FORM modules: 
({\bf{FF}}) {\it Form Factor},
({\bf{HA}}) {\it Helicity Amplitudes}, and ({\bf{BR}}) {\it Bremsstrahlung}.

Once the three FORM codes for the calculation of
({\bf{FF}}), ({\bf{HA}}) and ({\bf{BR}})
have been compiled and outputs transferred to the software package {\bf s2n.f},
one can get the numerical results. 
The user guide for running {\tt SANC} is given in Ref.~\cite{SANC1.00}.

We convolute the partonic sub-process cross section with quark density functions to get the 
cross section at the hadronic level. One must avoid double counting 
of the quark mass singularities,  subtracting them from the density functions. 
   
{\tt SANC} version {\tt v1.00} is accessible from servers at Dubna 
{\it http://sanc.jinr.ru/ (159.93.74.10)} and CERN 
{\it http://pcphsanc.cern.ch/ (137.138.180.42)}.
      
\section{QCD environment}

The QCD environment of {\tt SANC} is a set of FORM procedures relevant 
for QCD. The basic 
procedure is {\tt QCDAlgebra.prc}, which calculates color weights for 
process diagrams.
It uses some common relations for $T^a$ matrices and structure constants $f^{abc}$.
We have changed several intrinsic {\tt SANC} procedures like {\tt FeynmanRules.prc},
{\tt MakeAmpSquare.prc}, {\tt Trace.prc} by including 
gluon-quark and gluon-gluon vertices in the {\tt FeynmanRules.prc} and giving
 color indices for quark bispinors. Quarks and leptons bispinors have different 
representations in {\tt SANC}.

For example, the Born amplitude of the Drell--Yan process with a charged current
taken from procedure {\tt VirtCC4fQCD.prc} is of the following form: 
\bqa
\rm   BornDYCC &=&  \frac{1}{8} \,i\, 
\frac{g^2}{s-\mwt}
\nll
   &\times& \rm Vb(ii,p1,h1,cl1) \,\, \gamma(ii,mu) \,\, \gamma6(ii) \,\, U(ii,p2,h2,cl2) 
                                             \,  \, \delta(cl1,cl2)\nll
   &\times& \rm Ub(jj,p3,h3) \,\, \gamma(jj,mu) \,\, \gamma6(jj) \, \, V(jj,p4,h4).
\eqa
Here $\rm U(ii,p2,h2,cl2)$ and $\rm Vb(ii,p1,h1,cl1)$ are bispinors of the incoming 
up quark  and down antiquark, 
$\rm cl1$ and $\rm cl2$ are their color indices. While
$\rm Ub(jj,p3,h3)$ and $\rm V(jj,p4,h4)$  are bispinors of the outgoing lepton pair;
$\rm \gamma(ii,mu)$ and $\rm \gamma6(ii)=I(ii)+\gamma5(ii)$ are Dirac matrices. 
Here
\begin{equation}
\mwt = \mw^2-i\mw\Gamma_{\sss W},
\end{equation}
where $\mw$ is the mass
 and $\Gamma_{\sss W}$ is the width of the W boson.

Using this environment we build a set of precomputation files. The user can find it in the system.
To compute the quark self-energy, one follows the sequence:\\[1mm]
{\bf QCD $\to$ Precomputation $\to$ Self $\to$ Quark $\to$ Quark Self} \\[1mm] 
in the QCD tree of {\tt SANC}.
Precomputed quark self-energies are used by FORM programs which calculate
quark  counter terms:\\[1mm] 
{\bf QCD $\to$ Precomputation $\to$ Self $\to$ Quark $\to$ QuarkRenConst}.\\[1mm]
After we have  the  one-loop amplitude of a given QCD process free from 
ultraviolet divergences, we can obtain a virtual radiative correction using 
the universal procedure  {\tt MakeAmpSquare.prc} to calculate the modulus squared
of the amplitude.
  
 \section{QCD radiative correction to {\bf b2q} decays }   

\begin{figure}[!ht]
\[
\begin{picture}(480,90)(0,0)
%--
  \Photon(100,60)(100,86){3}{5}
    \Vertex(50,15){2}
    \Vertex(100,60){2}
    \Vertex(150,15){2}
  \ArrowLine(0,15)(50,15)
  \ArrowLine(50,15)(100,60)
  \ArrowLine(100,60)(150,15)
  \ArrowLine(150,15)(200,15)
  \Gluon(50,15)(150,15){2}{16}
\Text(4,0)[lb]  { $p_2$ } 
\Text(0,25)[lb]  { $q_2$ }   
\Text(100,0)[bc] { $g $  }
\Text(180,0)[lb]  { $p_1$}
\Text(186,25)[lb]  { $q_1$}
\Text(110,90)[bc] { $W ~or~ Z  ~or~ H$  }
%--
    \Vertex(350,15){2}
    \Vertex(430,15){2}
  \ArrowLine(300,15)(350,15)
  \ArrowLine(350,15)(430,15)
  \ArrowLine(430,15)(480,15)
  \GlueArc(390,15)(40,0,180){2}{20}
\Text(400,60)[bc] { $g $  }
\Text(300,25)[lb]  { $q$ }  
\Text(460,25)[lb]  { $q$ }
\Text(400,0)[bc]  { $q$ } 
%--
\end{picture}
\]
\vspace*{-6mm}
\caption {QCD vertex with two quark legs 
 and self energy for quarks.}
\label{Oneloop}
\end{figure}
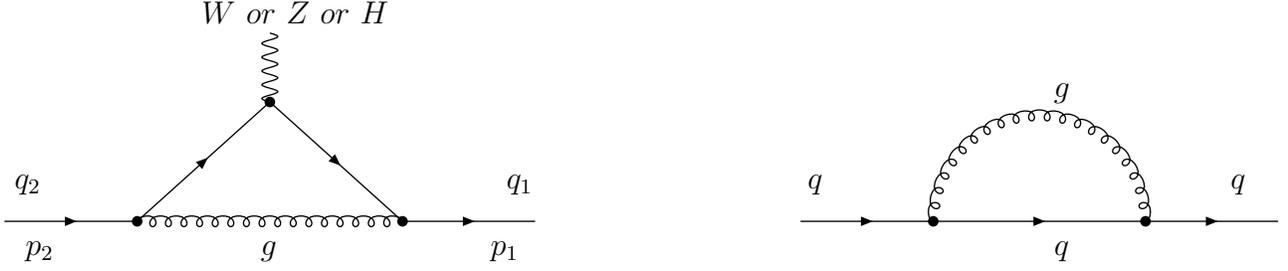

We started to work with simple processes of boson decays
suitable to test the QCD environment of {\tt SANC}. One-loop Feynman diagrams 
(see Fig.~\ref{Oneloop}) of these processes
contain only a vertex with one virtual gluon and two quark legs, and 
corresponding QCD counter terms.

Following the standard procedure of {\tt SANC}, we calculate the virtual part of QCD
corrections to these boson decays. 

\begin{figure}[!ht]
\[
\begin{picture}(490,80)(0,0)
%--
  \Photon(140,30)(140,66){3}{6}
    \Vertex(50,30){2}
    \Vertex(140,30){2}
  \ArrowLine(0,30)(50,30)
  \ArrowLine(50,30)(140,30)
  \ArrowLine(140,30)(190,30)
  \Gluon(50,30)(80,0){2}{10}
\Text(4,15)[lb]  { $p_2$ } 
\Text(0,35)[lb]  { $q_2$ }   
\Text(56,4)[lt] { $p_4 $  }
\Text(90,0)[bc] { $g $  }
\Text(90,35)[bc]{$p_2-p_4$ } 
\Text(176,15)[lb]  { $p_1$}
\Text(180,35)[lb]  { $q_1$}
\Text(145,65)[lt]  { $p_3$}
\Text(140,70)[bc] { $W ~or~ Z  ~or~ H$  }
%--
  \Photon(300,30)(300,66){3}{6}
    \Vertex(300,30){2}
    \Vertex(380,30){2}
  \ArrowLine(250,30)(300,30)
  \ArrowLine(300,30)(390,30)
  \ArrowLine(380,30)(440,30)
  \Gluon(380,30)(420,0){2}{10} 
\Text(254,15)[lb]  { $p_2$ } 
\Text(250,35)[lb]  { $q_2$ } 
\Text(396,4)[lt] { $p_4 $  }
\Text(425,0)[lt] { $g $  }
\Text(340,35)[bc]{$p_1+p_4$ } 
\Text(426,15)[lb]  { $p_1$}
\Text(430,35)[lb]  { $q_1$}
\Text(305,65)[lt]  { $p_3$}
\Text(300,70)[bc] { $W ~or~ Z  ~or~ H$  }
%--
\end{picture}
\]
\vspace*{-6mm}
\caption {Gluon bremsstrahlung from  two quark legs.}
\label{BqqBrem}
\end{figure}
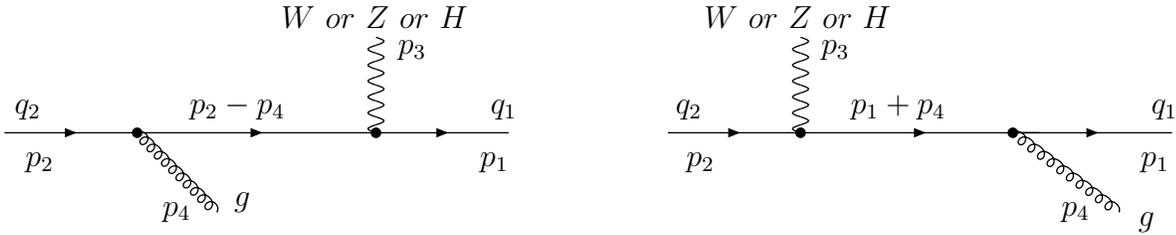

The real part of QCD corrections is created by gluon emission from every
quark leg (see Fig.~\ref{BqqBrem}). 

Gluon bremsstrahlung implementation needs another set of procedures, 
specific  for every process. Procedures {\tt BremWqqQCD.prc},
{\tt BremZqqQCD.prc}, {\tt BremHqqQCD.prc} prepare the 
corresponding amplitudes for $W \to qq $, $Z \to qq $, $H \to qq $ processes .
 Procedures {\tt SoftWqqQCD.prc}, {\tt SoftZqqQCD.prc},
{\tt SoftHqqQCD.prc} calculate
 the soft gluon  bremsstrahlung contributions analytically. Procedures {\tt HardWqqQCD.prc},
 {\tt HardZqqQCD.prc}, {\tt HardHqqQCD.prc}
prepare  fully differential expressions of hard  gluon brems\-strahlung 
contributions.
These expressions may be used by Monte Carlo generators (or integrators).
The {\it Hard} procedures continue the analytical calculation
of hard gluon contributions by integrating over the angle between one of the quarks
and the gluon and over the 
invariant mass of the final particles.  

Here we present QCD corrections
to  boson decays. Analytical expressions are too cumbersome 
to be presented in this paper.  
One can access these results in the system. Here we cast the corrected decays width in the form
\bqa
{\Gamma^{\rm 1loop}}=
   \Gamma^{\rm Born} \,\left[ 1 + \delta(m_b,m_{q_1},m_{q_2}) \right],
\label{deltaB}
\eqa
where $m_b$, $m_{q_1}$ and $m_{q_2}$ are the masses of the boson and two quarks and $\delta$ is the correction.

In regard to vector boson decays everything is straightforward.
In the Table~\ref{deltaVqq} we give one-loop numerical 
results for the function
$\delta(m_V,m_{q_1},m_{q_2})$ in percent, 
where $m_V$ is the mass of the vector bosons.
  
\begin{table}
\begin{center}
\large 
\begin{tabular}{|l|l|l|}
\hline 
process      & $\delta(m_b,m_{q1},m_{q2})$ & $\%$  \\ [3mm]\hline \hline
$ W \to c\bar{s} $& $\delta(m_W,m_c,m_s)$  & +3.44  \\ \hline 
$ Z \to b\bar{b} $& $\delta(m_Z,m_b,m_b)$  & +3.88  \\ \hline
$ Z \to u\bar{u} $& $\delta(m_Z,m_u,m_u)$  & +3.41  \\ \hline
\end{tabular}
\end{center}
\caption{Function $\delta(m_V,m_{q1},m_{q2})$ in percent.}
\label{deltaVqq}
\end{table}

The well known  formula for vector boson decay into massless quarks is:
\bqa
\lim\limits_{m_q\to 0}\delta(m_H,m_q,m_q)  = \frac{3}{4}\, C_f\, \frac{\als}{\pi}, 
\qquad
C_f=\frac{4}{3},
\eqa
which  gives $3.41 \%$ for all decays of vector bosons. 
One can see that our numbers are in a good agreement with the classic result.
However, mass effects are significant.
  
QCD radiative corrections to the Higgs boson decay into a quark pair have been 
considered earlier by one of the authors in Ref.~\cite{BaVilPXX}. 
There the one-loop 
QCD correction was presented keeping the masses of the outgoing quarks (Eq.~4.1)
and neglecting these masses everywhere except in logarithms (Eq.~4.3). 
In the framework of {\tt SANC} we reproduced these results. 
Here we give  the function $\delta(m_H,m_{q_1},m_{q_2})$ of the QCD 
correction in (\ref{deltaB}) 
for the Higgs boson decay keeping the masses of the outgoing quarks
in logarithms only: 
\bqa
\lim\limits_{m_q\to 0}\delta(m_H,m_q,m_q) =
 C_f\, \frac{\als}{\pi} \, 
\left[ \frac{9}{4}+ \frac{3}{2} \ln  \left(\frac{m^2_q}{M^2_H}\right) \right].
\eqa
    The term with the
large logarithm,
\bqa
 C_f\, \frac{\als}{\pi} \,\left[ \frac{3}{2} 
          \ln\left(\frac{m^2_q}{M^2_H}\right) \right],
\eqa
was obtained first and discussed in Ref.~\cite{BraLev}. It doesn't violate the
Kinoshita-Lee-Nauenberg theorem since the Higss-quark coupling constant 
is proportional to the quark mass. Moreover, resummation of these large logarithms
in all orders of the perturbation theory is possible as suggested in 
Ref.~\cite{BraLev}.
 
\section{QCD radiative corrections to semi-leptonic top quark decay}

Here we discuss in detail the analytic calculation of QCD radiative corrections
 to the semi-leptonic mode of the top quark decay  $t \to b \ell^+ \nu_\ell $.

First we consider the 
easier  ``cascade'' calculation of the top decay: 
\begin{center}
\begin{picture}(120,30)(0,0)
\Text(0,20)[lb]  { $t \longrightarrow  b \, + \, W^+$}
\Line(63,17)(63,5)
\LongArrow(63,5)(90,5)
\Text(100,0)[lb] { $\ell^+  \,+ \, \nu_\ell $}    
\end{picture}
\end{center}

 We use the formula for the process with
production of an unstable particle (here it is W boson) given in the 
book ~\cite{Bilenky}
to treat the top decay partial width:
\bqa 
\Gamma_{ t \to b \ell^+ \nu_\ell }= 
\frac{\Gamma_{ t \to b W^+}\,\, \Gamma_{ W^+ \to \ell^+ \nu_\ell}}
                    {\Gamma_{W}},
\label{Gcascade}
\eqa
where $\Gamma_{W}$ is total width and $\Gamma_{ W^+ \to \ell^+ \nu_\ell}$ 
is partial width of the W boson. 
The formula (\ref{Gcascade})
is valid when the total width of the unstable particle (i.e. W boson) is 
much less then  the mass of this particle. The order of magnitude estimate 
of the relative precision for  
this formula is ~${\displaystyle \Gamma_{W}/\mw }$. 

There are not any QCD corrections to the partial width 
$\Gamma_{ W^+ \to \ell^+ \nu_\ell}$ of the W boson. So we take here this 
partial width in Born approximation.
The QCD corrections are
present only in the partial width $\Gamma_{ t \to b W^+}$ of the top quark.

\subsection{QCD radiative correction to decay  $t \to b W^+$}

This mode of the top decay is treated in {\tt SANC}
in the same way as the {\bf b2q} decays
considered in the previous section.  
The virtual part of QCD
corrections to the decay $t \to b W^+$ is  similar to that of
the decay $W^- \to d  {\bar u}$. One can prepare it by
the procedure {\tt VirtTopWbQCD.prc} and then
calculate the gluon bremsstrahlung contribution using procedures 
specific  for decay $t \to b W^+$: {\tt BremTopWbQCD.prc}, {\tt SoftTopWbQCD.prc},
 {\tt HardTopWbQCD.prc}. In result we obtain:
\bqa
\Gamma_{ t \to b W^+} =\Gamma_{ t \to b \ell^+ \nu_\ell }^{\rm Born} \,(1+\delta),
\eqa
where $\delta$ is the QCD correction  to the decay $t \to b W^+ $ .

The authors of  the Ref.~\cite{Arbuzov2006}  have studied the 
effects of the total width
${\Gamma_{W}}$  in cascade calculations of EW corrections to the top quark decay
due to the photon emittion from the W boson.
However here we deal with gluons, they  are not 
emitted from  the  W boson, so the QCD correction to the 
decay $t \to b W^+$ does not depend on ${\Gamma_{W}}$. Results
of the narrow width cascade approximation  for QCD corrections to
the top quark decay  $ t \to b e^+ \nu_e $ are given in Table~\ref{cascade}. 

{\large 
\begin{table}
\begin{center}
\begin{tabular}{|l|l|l|l|}
\hline 
decay       & $t\to b W^+$&$W^+\to e^+ \nu_e$&$t\to b e^+ \nu_e$\\
             &              &                      & cascade \\ [3mm]\hline \hline
$\Gamma^{\rm Born}$  &    1.5930     &   0.22018       & 0.16405 \\ \hline 
$\Gamma^{\rm 1loop}$& 1.4764 &  0.22018   & 0.15204 \\ \hline
$\delta, \% $        &  - 7.32    &   0            & -7.32  \\ \hline
\end{tabular}
\end{center}
\caption{Cascade approximation  for QCD corrections to
the top quark decay  $t \to b e^+ \nu_e $.}
\label{cascade}
\end{table}
}
We see that the QCD correction to  $ \Gamma_{ t \to b \ell^+ \nu_\ell }$
is the same as the QCD correction 
to  $\Gamma_{ t \to b W^+}$. This is so because
according to the formula~(\ref{Gcascade}) we have
\bqa 
\Gamma_{ t \to b  \ell^+ \nu_\ell }= 
\frac{\Gamma_{ t \to b W^+}^{\rm Born}\,(1+\delta)\, 
\Gamma_{ W^+ \to \ell^+ \nu_\ell}^{\rm Born}} {\Gamma_{W}} 
 =\Gamma_{ t \to b \ell^+ \nu_\ell }^{\rm Born} \,(1+\delta).
\label{deltaCascade}
\eqa

\subsection{One-loop QCD amplitude of decay $t \to b + \ell^+ +\nu_\ell$}

The one-loop QCD  amplitude comes from the following gauge independent set
of diagrams:

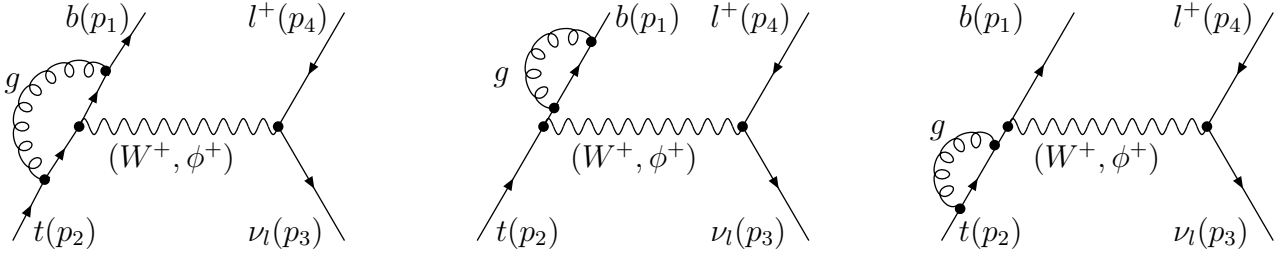
\begin{figure}[!ht]
\[
\begin{picture}(475,86)(0,0)
%--
  \Photon(25,43)(100,43){3}{10}
    \Vertex(100,43){2}
  \ArrowLine(125,86)(100,43)
  \ArrowLine(100,43)(125,0)
%--
  \ArrowLine(0,0)(12,23)
  \Vertex(12,23){2}
  \ArrowLine(12,23)(25,43)
    \Vertex(25,43){2}
  \ArrowLine(25,43)(37,66)
  \Vertex(35,64){2}
  \ArrowLine(37,66)(50,86)
  \GlueArc(25,43)(22,60,240){3}{8}
%--
\Text(16,78)[lb]  { $b(p_1)$}
\Text(4,9)[lt]  { $t(p_2)$    }
\Text(60,25)[bc] { $(W^+,\phi^+)$  }
\Text(0,56)[bc] { $g $  }
\Text(85,78)[lb]{ $l^+(p_4)$}
\Text(85,9)[lt]{ $\nu_l(p_3)$      }
%--
  \Photon(200,43)(275,43){3}{10}
    \Vertex(275,43){2}
  \ArrowLine(300,86)(275,43)
  \ArrowLine(275,43)(300,0)
%--
  \ArrowLine(175,0)(200,43)
    \Vertex(200,43){2}
  \ArrowLine(200,43)(225,86)
  \GlueArc(209,64)(13,45,255){3}{5}
  \Vertex(218,75){2}
  \Vertex(204,50){2}
%  \GlueArc(190,63)(13,120,300){3}{5}
%--
\Text(224,78)[lb]  { $b(p_1)$}
\Text(179,9)[lt]  { $t(p_2)$    }
\Text(235,25)[bc] { $(W^+,\phi^+)$  }
\Text(185,58)[bc] { $g$ }
\Text(260,78)[lb]{ $l^+(p_4)$}
\Text(260,9)[lt]{ $\nu_l(p_3)$      }
%--
  \Photon(375,43)(450,43){3}{10}
    \Vertex(450,43){2}
  \ArrowLine(475,86)(450,43)
  \ArrowLine(450,43)(475,0)
%--
  \ArrowLine(350,0)(375,43)
    \Vertex(375,43){2}
  \ArrowLine(375,43)(400,86)
  \GlueArc(363,25)(13,60,240){3}{5}
  \Vertex(370,36){2}
  \Vertex(357,12){2}
%--
\Text(354,78)[lb]  { $b(p_1)$}
\Text(354,9)[lt]  { $t(p_2)$    }
\Text(410,25)[bc] { $(W^+,\phi^+)$  }
\Text(350,38)[bc] { $g$  }
\Text(435,78)[lb]{ $l^+(p_4)$}
\Text(435,9)[lt]{ $\nu_l(p_3)$      }
%--
\end{picture}
\]
\vspace*{-6mm}

\caption[$f_1\bar{f}_1\to f\bar{f}$ process] 
        {One-loop QCD diagrams of t quark decay: vertex and two counter terms.}
\label{tblnOneloop}
\end{figure}

\noindent
The QCD part of the amplitude  has the same structure as the electroweak one
(see Ref.~\cite{SANC1.00}):
\bqa
{\cal A} = V_{tb}\,\frac{g^2}{8}\, {{\bar U}_b(p_1)} \,\Big[ &&
+  i \, \gamma_{\mu} \left(1+\gamma_5 \right) {\cal F}^{\rm QCD}_{\sss{LL}}(s)
+ i \, \gamma_{\mu} \left(1-\gamma_5 \right) {\cal F}^{\rm QCD}_{\sss{RL}}(s) \nll
&&+ D_\mu \left( 1+\gamma_5 \right)  {\cal F}^{\rm QCD}_{\sss{LD}}(s)\,
  +  D_\mu \left( 1-\gamma_5 \right) {\cal F}^{\rm QCD}_{\sss{RD}}(s) \,\Big] \,
               {U_t(p_2)}                   \, \nll
 &&\times  \frac{1}{s-\mwt} \,\, {{\bar U}_{\nu_l}(p_3)}\,
    \gamma_\mu \left( 1+\gamma_5 \right) \,   {V_l(p_4)}.
\label{A}
\eqa
%---
Here $D_\mu$ and 4-momentum conservation read
\bq
D_\mu=(p_1+p_2)_\mu\,,  \quad p_2=p_1+p_3+p_4\,,
\eq
%--
the invariant $s$  is
\bq
s=-(p_2-p_1)^2,
\eq
%--
$V_{tb}$ is the element of the CKM matrix; ${\bar U }$, $U$ and $V$ 
are the corresponding bispinors.

Form factors  obtained by  FORM code {\bf{FF}} (see Fig.~\ref{QCDtree}) are already 
free from ultraviolet divergences:
\bqa
{\cal F}^{\rm QCD}_{\sss{LL}}(s)&=&{C_f} \frac{\als}{4 \pi} \, \Bigg[
    \ln\left(\frac{\mtp^2}{\mg^2}\right) + \ln\left(\frac{\mbt^2}{\mg^2}\right)
      - 2\, (\mtp^2 + \mbt^2 - s)\, C_0(-\mtp^2,-\mbt^2, - s;\mtp,\mg,\mbt) \nll
&& +  \left(\frac{1}{\beta(\mtp^2,\mbt^2,s)}
            -\frac{\sqrt{\lambda(s,\mtp^2,\mbt^2)}}{2 s}\right)
  L\left(m_t^2,m_b^2,s\right) 
    - 4     - \frac{\mtp^2-\mbt^2}{2 s} \, \ln\left(\frac{\mbt^2}{\mtp^2}\right)
                    \Bigg],\nll
{\cal F}^{\rm QCD}_{\sss{RL}}(s)&=& {C_f} \frac{\als}{2 \pi} \, 
              \frac{\mtp \mbt}{\sqrt{\lambda(s,\mtp^2,\mbt^2)}} \,
 L\left(m_t^2,m_b^2,s\right),\nll
{\cal F}^{\rm QCD}_{\sss{RD}}(s)&=& - {C_f} \frac{\als}{8 \pi} \, 
   \frac{\mtp}{s} \, \Bigg[  \ln\left(\frac{\mbt^2}{\mtp^2}\right)
   + \frac{1}{\beta(\mtp^2,-s,\mbt^2)}
    L\left(m_t^2,m_b^2,s\right)\Bigg],\nll
{\cal F}^{\rm QCD}_{\sss{LD}}(s)&=&{C_f} \frac{\als}{8 \pi} \, 
   \frac{\mbt}{s} \, \Bigg[ \ln\left(\frac{\mbt^2}{\mtp^2}\right)
   + \frac{1}{\beta(\mtp^2,s,\mbt^2)}
 L\left(m_t^2,m_b^2,s\right)\Bigg],
\eqa 
where  
\bqa
\lambda(s,\mtp^2,\mbt^2) = s^2+\mtp^4+\mbt^4-2s\mtp^2-2s\mbt^2-2\mtp^2\mbt^2,\nll 
\beta(\mtp^2,\mbt^2,s)=\frac{\sqrt{\lambda(s,\mtp^2,\mbt^2)}}{\mtp^2+\mbt^2-s},
\qquad
\beta(\mtp^2,s,\mbt^2)=\frac{\sqrt{\lambda(s,\mtp^2,\mbt^2)}}{\mtp^2+s-\mbt^2},\nll
\beta(\mtp^2,-s,\mbt^2)=\frac{\sqrt{\lambda(s,\mtp^2,\mbt^2)}}{\mtp^2-s-\mbt^2},
\qquad
L\left(x,y,z\right) = \ln \frac{1+\beta(x,y,z)}{1-\beta(x,y,z)}.
\eqa

The gluon infrared singularity is regularized by a fictitious 
gluon mass  $ \mg$. The
 Passarino-Veltman function 
$ C_0(-\mtp^2,-\mbt^2,-s;\mtp,\mg,\mbt)$
is infrared singular. The lepton mass is neglected.

The helicity amplitudes for this mode of top quark decay are the same as those
 given in~\cite{SANC1.00} but we have to take QCD form factors 
${\cal F}^{\rm QCD}_{\sss{LL}}(s)$, ${\cal F}^{\rm QCD}_{\sss{RL}}(s)$,  
${\cal F}^{\rm QCD}_{\sss{LD}}(s)$ and  ${\cal F}^{\rm QCD}_{\sss{RD}}(s)$ 
instead of EW ones.

\subsection{Virtual  QCD correction}

The three particle phase space element is

\bqa
{ d \Phi^{(3)}}= 
   \frac{d s}{2 \pi} \,\, \frac{\sqrt{\lambda(s,\mtp^2,\mbt^2)}}{8 \pi \mtp^2} 
    \quad \frac{d \cos\thle}{16 \pi} \,=\, \frac{d s \,\, d u}{128 \pi^3 \, \mtp^2},
\eqa
where the invariant $u$ is related to the angle  $\thle$ between the lepton 
momentum $ \vec{p_4}$ in the R-frame ($\vec{p}_3+\vec{p}_4=0$)
and the momentum $\vec{p_1}$ of the $b$ quark, as follows:
\bq
   u = \frac{1}{2} \left[\mtp^2+\mbt^2-s 
                       + \sqrt{\lambda(s,\mtp^2,\mbt^2)} \cos\thle\right].
\eq

The angle  $\thle$ varies from 
$0$ to $\pi$ and the invariant $s$ varies in the interval
\bq
         m_{\ell}^2 \, \leq \, s \,  \leq \, (\mtp - \mbt)^2.
\label{sTop}
\eq  

Using the formula
\bq
d \Gamma =  \frac{1}{2 \mtp}\, {\mid V_{tb}\mid}^2 
               {\mid {\cal A}\mid}^2\, { d \Phi^{(3)}},  
\label{DifA0^2}
\eq
we obtain by the procedure {\tt VirtTop3fQCD.prc} the
virtual QCD correction to the differential width of the top quark decay
 ${\displaystyle \frac{d^2 \Gamma_{\rm Virt}(s,u)}{d s \,d u}}$ . Here 
we give this  analytic expression in the form:
\bqa 
 \frac{d^2 \Gamma_{\rm Virt}(s,u)}{d s \,d u}&=&
 \frac{d^2 \Gamma_0(s,u)}{d s \,d u} \,\,\frac{\alpha_s}{2 \pi}
  \,\, \left[       {\cal F}^{\rm QCD}_{\sss{LL}}(s)
            -\mtp {\cal F}^{\rm QCD}_{\sss{RD}}(s)              
            -\mbt {\cal F}^{\rm QCD}_{\sss{LD}}(s) \right] \nll
&+&    {\mid V_{tb}\mid}^2 \frac{{G^2_{\sss F}}}{16 \pi^3 \mtp^3} \,\,
       \frac{\mw^4 \, s}{{\mid {s-\mwt}\mid}^2} \,\,\frac{\alpha_s}{2 \pi}\\
&& \qquad \left[  -\mtp \,\mbt\,  {\cal F}^{\rm QCD}_{\sss{RL}}(s)
            +\mtp \, u \,   {\cal F}^{\rm QCD}_{\sss{RD}}(s)              
            +\mbt \, u \,   {\cal F}^{\rm QCD}_{\sss{LD}}(s) \right], \nonumber
\label{Virt}
\eqa
where  
\bqa
\frac{d^2 \Gamma_0(s,u)}{d s \,d u}  =
     {\mid V_{tb}\mid}^2 \frac{{G^2_{\sss F}}}{16 \pi^3 \mtp^3} \,\,
       \frac{\mw^4 }{{\mid {s-\mwt}\mid}^2} \,\,
        \left(\mtp^2-u\right) \,   \left(u - \mbt^2\right) .
\eqa
is the differential width of the top quark decay in
the Born approximation. 

\subsection{Gluon bremsstrahlung  corrections} 
 The gluon bremsstrahlung amplitude is prepared by the procedure 
{\tt BremTop3fQCD.prc}.  
  
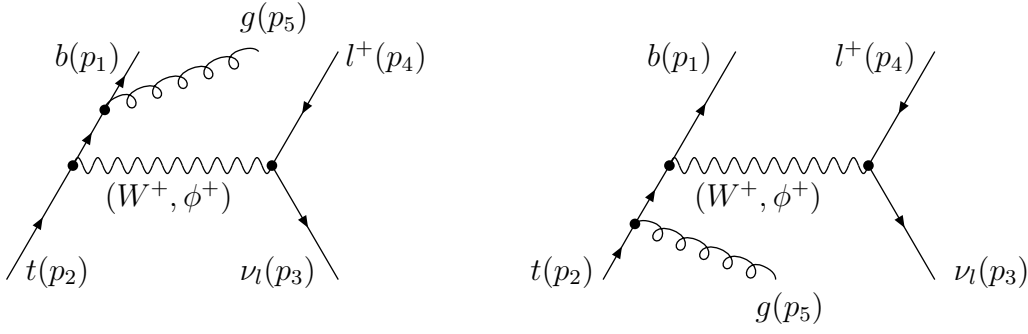
\begin{figure}[!ht]
\[
\hspace{20mm}
\begin{picture}(475,86)(200,0)
%--
  \Photon(200,43)(275,43){3}{10}
    \Vertex(275,43){2}
  \ArrowLine(300,86)(275,43)
  \ArrowLine(275,43)(300,0)
%--
  \ArrowLine(175,0)(200,43)
    \Vertex(200,43){2}
  \ArrowLine(200,43)(212,64)
    \Vertex(212,64){2}
  \ArrowLine(212,64)(225,86)
    \Gluon(212,64)(270,86){3}{5}
%--
\Text(190,78)[lb]  { $b(p_1)$}
\Text(179,9)[lt]  { $t(p_2)$}
\Text(235,25)[bc] { $(W^+,\phi^+)$}
\Text(300,78)[lb]{ $l^+(p_4)$}
\Text(260,9)[lt]{ $\nu_l(p_3)$}
\Text(260,105)[lt]{ $g(p_5)$}
%--
\SetOffset(50,0)
  \Photon(375,43)(450,43){3}{10}
    \Vertex(450,43){2}
  \ArrowLine(475,86)(450,43)
  \ArrowLine(450,43)(475,0)
%--
  \ArrowLine(350,0)(362,21)
    \Vertex(362,21){2}
  \ArrowLine(362,21)(375,43)
    \Vertex(375,43){2}
  \ArrowLine(375,43)(400,86)
  \Gluon(362,21)(415,0){3}{5}
%--
\Text(364,78)[lb]  { $b(p_1)$}
\Text(320,9)[lt]  { $t(p_2)$    }
\Text(410,25)[bc] { $(W^+,\phi^+)$  }
\Text(435,78)[lb]{ $l^+(p_4)$}
\Text(480,9)[lt]{ $\nu_l(p_3)$      }
\Text(405,-5)[lt]{ $g(p_5)$}
%--
\end{picture}
\]
\caption[Gluon bremsstrahlung diagrams of t quark decay.] 
        {Gluon bremsstrahlung diagrams of t quark decay.}
\label{tblnBrem}
\end{figure}

The conservation of 4-momentum reads
\bq
p_2=p_1+p_3+p_4 + p_5 .
 \eq

The energy of the emitted gluon is 
obtained in the rest system
${\bar p_3}+{\bar p_4} + {\bar p_5} = 0$:
\bq
   p^0_5=\frac{s-s'}{2 \,\sqrt{s}}
\label{Egluon}
\eq
where ${ s'}$ is the invariant mass of the two final leptons,
$s'=-(p_3+p_4)^2$. ~It varies in the interval :
\bq
     m_{\ell}^2 \, \leq \, s' \, \leq \, s . 
\label{s'Top}  
\eq
We see from (~\ref{Egluon}), (\ref{sTop}) and (\ref{s'Top}) that the energy of 
the emitted gluon could go to zero even when the invariant $s$
goes to its minimal value $m_{\ell}^2 $.  

An auxiliary parameter $\bar\omega$ separates
the soft and hard  gluon contributions.
Gluon energy for the soft gluon bremsstrahlung lies in the limits
\bqa
    0 \leq  p^0_5 \leq \bar\omega,
 \eqa
where $\bar\omega $ is arbitrary small. Therefore the soft 
bremsstrahlung amplitude is factorized  by the Born amplitude
and  the kinematics of the soft bremsstrahlung  is Born-like. The 
four-particle phase space element in this case  is a product of the same 
three-particle phase space element $d \Phi^{(3)}$ and 
the phase space element of the emitted soft gluon.

By the procedure {\tt SoftTop3fQCD.prc} the soft 
gluon contribution is obtained in the form:
\bqa
 \frac{d^2 \Gamma_{\rm Soft}(s,u)}{d s \,d u}=
\frac{d^2 \Gamma_0(s,u)}{d s \,d u} \, {C_f} \,\frac{\alpha_s}{\pi}\,
                                            \,{\delta}_{\rm Soft}(s),
\eqa
where the integration in 
\bqa
{\delta}_{\rm Soft}(s)= \frac{1}{4} \int\limits_{-1}^{+1} d \xi \,
  \left[ \ln\left(\frac{{\bar\omega}^2}{\mg^2}\right)
  + \ln\left(1 - {\xi}^2  \right) \right] \,
   \left[ \frac{p_1^2}{(p_1 \cdot n)^2} 
           - \frac{2 p_1 \cdot p_2}{(p_1 \cdot n) (p_2 \cdot n)}
           + \frac{p_2^2}{(p_2 \cdot n)^2}  \right]
\eqa
over the angle of the gluon emission is performed in the same rest system
${\bar p_3}+{\bar p_4} + {\bar p_5} = 0 = {\bar p_2} -{\bar p_1} $. 
Taken the gluon momentum
$p_5 $  as $p_5 = n p^0_5$
here we use the unit 4-vector $n=(1,\bar n)$  where $\bar n$      
has the direction of the  vector ${\bar p_5}$ . The standard 
method~\cite{'tHooft:1978xw} by t'Hooft and Veltman was applied.

    The infrared divergences here are the same as the ones in the virtual 
gluon contribution 
${\displaystyle \frac{d^2 \Gamma_{\rm Virt}(s,u)}{d s \,d u} }$
but with the opposite sign. So, the sum of virtual and soft gluon contributions 
does not contain any infrared divergences. We give here the 
expression of this sum integrated by {\tt IntcTop3fQCD.prc} over the invariant $u$
 (i.e. over the angle  $\thle$): 

\bqa
\frac{d \Gamma_{\rm Virt}(s)}{d s}
 &+&\frac{d \Gamma_{\rm Soft}(s)}{d s}
 = 
\frac{d \Gamma_0(s)}{d s} \,\, \, {C_f} \,\frac{\alpha_s}{2 \pi}\, 
   \Biggl\{ \,{\ln}\left(\frac{4\,{\bar\omega}^2}{s}\right) \left[ 
        \frac{1}{\beta\left(\mtp^2,\mbt^2,s\right)}L(\mtp^2,\mbt^2,s) -2  \right] \nll
  &&+ {\ln}\left(\frac{m_t^2}{s}\right) \, \left[ 
       \frac{1}{\beta(m_t^2,m_b^2,s)} \,  L(m_t^2,s,m_b^2)
        +\frac{m_t^2}{2 s}-\frac{m_b^2}{2 s}+ 1 \right] \nll
  &&- {\ln}\left(\frac{m_b^2}{s}\right) \, \left[ 
       \frac{1}{\beta(m_t^2,m_b^2,s)} \,  L(m_t^2,-s,m_b^2)
        + \frac{m_t^2}{2 s} -\frac{m_b^2}{2 s}  - 1 \right] \nll 
  &&+ \frac{1}{\beta(m_t^2,m_b^2,s)} \left[ L^2(m_t^2,s,m_b^2) + 4\,
 {\rm Li_2}\left(\frac{2 \beta(m_t^2,s,m_b^2)}{1 + \beta(m_t^2,s,m_b^2)}\right)\right]\nll
  &&- \frac{1}{\beta(m_t^2,m_b^2,s)}\left[  L^2(m_t^2,-s,m_b^2) + 4\,
 {\rm Li_2}\left(\frac{2 \beta(m_t^2,-s,m_b^2)}{1 + \beta(m_t^2, - s,m_b^2)}\right)
                                               \right]  \nll 
  &&+ \frac{1}{\beta(m_t^2,m_b^2,s)}  \, L(m_t^2,m_b^2,s)
      + \frac{1}{\beta(m_t^2, - s,m_b^2)} \, L(m_t^2,-s,m_b^2) \nll
  &&+ \frac{1}{\beta(m_t^2,s,m_b^2)}  \, L(m_t^2,s,m_b^2) -4
 \nll 
   &&-  \, \left(\frac{\sqrt{\lambda(s,m_t^2,m_b^2)}}{2 s} 
                  + \frac{m_t m_b}{\sqrt{\lambda(s,m_t^2,m_b^2)}} \right) L(m_t^2,m_b^2,s)
 \Biggr\} \\
    &&- 
  {\mid V_{tb}\mid}^2 \frac{{G^2_{\sss F}}\,\mbt }{64 \pi^3 \mtp^2}\frac{{C_f} \als}{\pi}
    \frac{ s \left(s-\mtp^2-\mbt^2+4 \mtp \mbt\right) \mw^4}
         {{\mid {s-\mwt}\mid}^2}
            \,\,  L(\mtp^2,\mbt^2,s). \nonumber
\eqa

%%%%%%%
%\vspace{5mm}

The hard gluon contribution is produced by the procedure {\tt HardTop3fQCD.prc}.
The gluon energy $p^0_5 $  for the hard gluon bremsstrahlung varies in the 
interval 
\bqa
    \bar\omega \, \leq \, p^0_5 \, \leq \, max(p^0_5).
 \eqa
The four particle phase space element is
\bqa
&&{ d \Phi^{(4)}}=
   \frac{d s}{128 \pi^3 \, \mtp^2}
      \,\, \frac{\sqrt{\lambda(s,\mtp^2,\mbt^2)}}{128 \pi^3 s}
      \,\, \left( s - s'\right) \, d s' \,
              d \cos\vartheta_1 \,d \cos\vartheta_4 d \varphi_4.
\eqa
For the hard gluon contribution the invariant
${ s'}$ varies in the interval:
\bq
     m_{\ell}^2 \, \leq \, s' \, \leq \, {s - 2 \bar\omega \,\sqrt{s}}.   
\eq
The separating parameter $\bar\omega $ is arbitrary small. The sum of soft 
and hard gluon contribution to the decay width has no trace of it.   

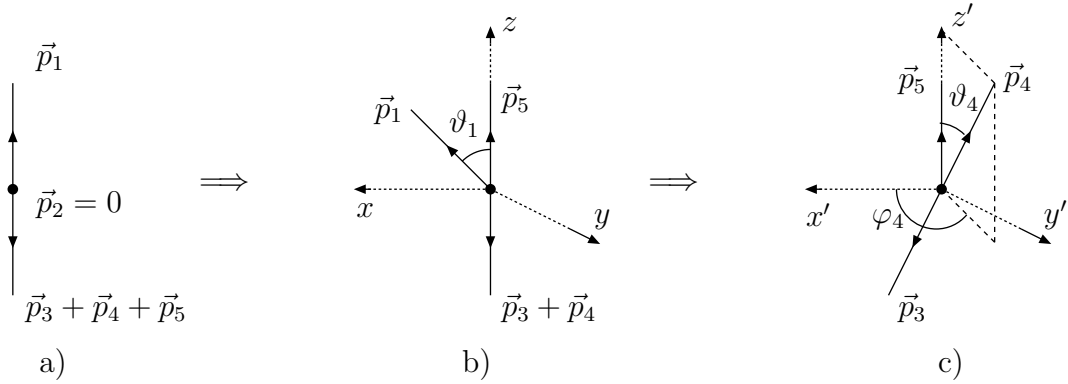
\begin{figure}[!ht]
\[
\begin{picture}(400,120)(0,0)
%--
  \Vertex(0,60){2}
  \ArrowLine(0,60)(0,100)
  \ArrowLine(0,60)(0,20)
\Text(5,60)[lt]  { ${\vec p_2}=0$ }
\Text(5,105)[lb]  { ${\vec p_1}$}
\Text(5,10)[lb] {${\vec p_3}+{\vec p_4}+{\vec p_5}$}
\Text(10,0)[lt] {a)}
\Text(80,60)[bc] {$\Longrightarrow$}
%--
  \Vertex(180,60){2}
  \ArrowLine(180,60)(180,100)
  \DashLine(180,100)(180,115){1}
  \LongArrow(180,115)(180,120) 
  \DashLine(130,60)(180,60){1}
  \LongArrow(135,60)(130,60)
  \DashLine(180,60)(210,45){1}
  \LongArrow(210,45)(220,40)
  \ArrowLine(180,60)(150,90)
  \CArc(180,60)(15,90,135)
  \ArrowLine(180,60)(180,20)
\Text(185,120)[lb]{$z$}
\Text(130,55)[lt]{$x$}
\Text(220,45)[lb]{$y$}
\Text(185,100)[lt]{${\vec p_5}$}
\Text(133,85)[lb]  { ${\vec p_1}$}
\Text(165,80)[lb] {$\vartheta_1$}
\Text(185,10)[lb] {${\vec p_3}+{\vec p_4}$}
\Text(170,0)[lt] {b)}
\Text(250,60)[bc] {$\Longrightarrow$}
%--
  \Vertex(350,60){2}
  \ArrowLine(350,60)(350,100)
  \DashLine(350,100)(350,115){1}
  \LongArrow(350,115)(350,120) 
  \DashLine(300,60)(350,60){1}
  \LongArrow(305,60)(300,60)
  \DashLine(350,60)(380,45){1}
  \LongArrow(380,45)(390,40)
  \ArrowLine(350,60)(370,100)
  \CArc(348,70)(15,45,85)
  \CArc(348,60)(15,180,320)
  \ArrowLine(350,60)(330,20)
  \DashLine(350,120)(370,100){2}
  \DashLine(350,60)(370,40){2}
  \DashLine(370,100)(370,40){2}
\Text(355,122)[lb]{$z'$}
\Text(300,55)[lt]{$x'$}
\Text(390,45)[lb]{$y'$}
\Text(335,107)[lt]{${\vec p_5}$}
\Text(354,90)[lb] {$\vartheta_4$}
\Text(325,45)[lb] {$\varphi_4$}
\Text(375,97)[lb] {${\vec p_4}$}
\Text(335,10)[lb] {${\vec p_3}$}
\Text(350,0)[lt] {c)}
%--
\end{picture}
\]
\caption[Systems]{a) Lab. system; b) Rest system 
${\vec p_3}+{\vec p_4}+{\vec p_5}=0$;
    c) Rest system ${\vec p_3}+{\vec p_4}=0$.}
\label{tblnKinem}
\end{figure}

The kinematics and choice of variables to be integrated over are illustrated
in Fig.~\ref{tblnKinem}.
The angle
 $\vartheta_1$ is between the b quark momentum ($\vec{p_1}$) and the emitted gluon 
momentum ($\vec{p_5}$), the angle  $\vartheta_4$ is between charged lepton 
 momentum ($\vec{p_4}$) and emitted gluon momentum ($\vec{p_5}$).
The z-axis  is chosen  here along  the momentum $\vec{p_5}$ of the emitted gluon. 

Using a formula analogous to~(\ref{DifA0^2}) 
we obtain  
the fully differential hard gluon contribution 
to the top quark decay. After integration over angles we get
\bqa
 \frac{d^2 \Gamma_{\rm Hard}(s,s')}{d s \,d s'}&=&
\frac{d \Gamma_0(s)}{d s}   \,\frac{{C_f}\,\alpha_s}{\pi\,(s - s')}\, 
 \frac{{\mid {s-\mwt}\mid}^2}
      {{\mid {s'-\mwt}\mid}^2}
   \,
 \Biggl[\frac{1}{{\beta}(\mtp^2,\mbt^2,s)}\, L(\mtp^2,\mbt^2,s)  - 2 \Biggr]  \nll
&+&  {\mid V_{tb}\mid}^2 \frac{{G^2_{\sss F}}\, }{96 \pi^3 \mtp^3}
   \,\frac{{C_f}\,\alpha_s}{\pi}\,\, \frac{\mw^4}{{\mid {s'-\mwt}\mid}^2} 
   \, \nll \times \Biggl\{ &-& \left[ (\mtp^2+\mbt^2)^2-(\mtp^2+\mbt^2)
    \left(\frac{7}{2} s +\frac{3}{2} s'\right)
    + 2 s^2 + s s' + s'^2 \right]\, L(\mtp^2,\mbt^2,s) \nll
&+& \frac{2}{s}\,\sqrt{\lambda(s,\mtp^2,\mbt^2)}
       \left[ s (\mtp^2+\mbt^2)-(2 s^2 + s s' + s'^2)\right]
       \Biggr\} .
\eqa

 Neglecting  the mass of the charged
lepton we obtain the QCD radiative correction to the decay width:
\bqa 
\Gamma_{ t \to b  \ell^+ \nu_\ell }&=& 
 \Gamma^{\rm Born}_{ t \to b  \ell^+ \nu_\ell } \,(1+\delta),\nll
          \delta &=& - 8.48 \%.
\label{genuine}
\eqa
Our result is in agreement with the result of 
Ref.~\cite{Fischer2001} ($\delta = - 8.5\%$).

Comparing the narrow width cascade approximation result in 
Table~\ref{cascade}  with the complete result~(\ref{genuine}) 
we see that the QCD radiative correction to the decay width
obtained by the  narrow width formula~(\ref{deltaCascade}) is near but not equal
to the complete result.

\section{QCD radiative corrections to Drell--Yan processes}
Here we present the results for the corrections to the charged (CC) and neutral (NC)
current Drell--Yan processes, ~$u\bar d \to \ell^+ \nu_\ell$ and 
$q \bar q \to \ell^+ \ell^-$, respectively. All formulas below
are shown at the partonic level.

At first we give expressions for cross sections in the Born approximation:
\bqa
{\hat \sigma}^{\rm CC}_0(s) &=&{\mid V_{ud}\mid}^2 \frac{{G^2_{\sss F}}}{18 \pi} \,\,
       \frac{\mw^4 }{{\mid {s-\mwt}\mid}^2} \,\,
\left(
s -\frac{3}{2} m^2_\ell + \frac{m^6_\ell}{2 s^2} 
\right),
\nll
{\hat \sigma}^{\rm NC}_0(s)&=& \pi \, \alpha^2 \,\beta(s,m^2_\ell,m^2_\ell)
\left[
\frac{4}{9 s} \left(1-\frac{m^2_\ell}{s} \right) V_0(s) 
+\frac{4 m^2_\ell}{3 s^2} V_a(s)
\right],
\label{BornDY}
\eqa
where
$ s = - (p_1+p_2)^2$, $p_1$ and $p_2$ are 4-momenta of the initial quarks;
$\displaystyle \beta(s,m^2_\ell,m^2_\ell)=\sqrt{1-\frac{4 m^2_\ell}{s}}$; 
$m_\ell$ is the lepton mass. Here we denote
\bqa
V_0(s)&=& Q^2_q Q^2_\ell + Q_q Q_\ell \left[ \chi_{\sss Z}(s)
   +\chi^*_{\sss Z}(s)\right]
v_q v_\ell + {\mid \chi_{\sss Z}(s) \mid}^2
\left( v^2_q+{I^{(3)}_q}^2\right)\left( v^2_\ell+{I^{(3)}_\ell}^2\right),
\nll
V_a(s) &=&V_0(s)-2{\mid \chi_{\sss Z}(s)\mid}^2 
\left( v^2_q+{I^{(3)}_q}^2\right)\left({I^{(3)}_\ell}\right)^2,
\nll && v_q = I^{(3)}_q - 2 Q_q \sin^2\theta_W, \qquad  \quad
   v_\ell = I^{(3)}_\ell - 2 Q_\ell \sin^2\theta_W.
\eqa
The $Z/\gamma$ propagator ratio $\chi_{\sss Z}(s)$
with $s$--dependent (or constant) $Z$ width
is given in~\cite{PhysPaNuc}.  

The one-loop QCD amplitude of the charged current Drell--Yan process
 is similar to~(\ref{A}) given in the previous section and the corresponding
amplitude of the neutral current Drell--Yan process has another form. However,
we neglect the terms proportional to the masses of the initial quarks and
therefore the virtual QCD corrections of both processes calculated 
in corresponding procedures {\tt VirtCC4fQCD.prc} and {\tt VirtNC4fQCD.prc} are
proportional to corresponding Born cross sections.

The gluon bremsstrahlung amplitudes both for CC and NC processes are prepared by 
the procedure {\tt Brem4fQCD.prc}. Here we give the sum of soft and virtual gluon 
contributions which does not contain any infrared divergences.

\bqa
{\hat \sigma}^{\rm CC}_{\rm Virt} + {\hat \sigma}^{\rm CC}_{\rm Soft} &=& 
{\hat \sigma}^{\rm CC}_0(s) \, \, C_f \, \frac{\als}{2 \pi} \Bigg\{
\ln\left(\frac{4 {\bar \omega}^2}{s} \right)
\left[ \ln\left( \frac{s}{m^2_u}\right) + \ln\left( \frac{s}{m^2_d}\right)- 2\right]
\nll && \qquad \qquad \quad
+\frac{3}{2}\ln\left( \frac{s}{m^2_u}\right) 
+\frac{3}{2}\ln\left( \frac{s}{m^2_d}\right)
-4 - \frac{\pi^2}{3}
\Bigg\},
\\
{\hat \sigma}^{\rm NC}_{\rm Virt} + {\hat \sigma}^{\rm NC}_{\rm Soft} &=& 
{\hat \sigma}^{\rm NC}_0(s) \, \, C_f \, \frac{\als}{\pi} \Bigg\{
\ln\left(\frac{4 {\bar \omega}^2}{s} \right) 
\left[ \ln\left( \frac{s}{m^2_q}\right) - 1\right]
\nll && \hspace{4.5cm}
+ \frac{3}{2}\ln\left( \frac{s}{m^2_q}\right) -2 -\frac{\pi^2}{6}
\Bigg\} ,
\eqa
where $m_q$, $m_u$, $m_d$ are quark masses.
Hard and soft gluon bremsstrahlung contributions are calculated in procedures
{\tt SoftCC4fQCD.prc}, {\tt HardCC4fQCD.prc} for CC and {\tt SoftNC4fQCD.prc}, 
{\tt HardNC4fQCD.prc}  for NC, respectively. We present here expressions 
for hard brem\-sstrahlung with extracted splitting function:
\bqa
\frac{d {\hat \sigma}^{\rm CC}_{\rm Hard}}{d s'} &=& 
{\hat \sigma}^{\rm CC}_0(s') \,\, C_f\,\frac{\als}{2 \pi} \frac{1}{s^2}
\frac{s^2+s'^2}{s-s'}\,
\left[ \ln\left( \frac{s}{m^2_u}\right)+\ln\left( \frac{s}{m^2_d}\right)- 2\right], 
\\
\frac{d {\hat \sigma}^{\rm NC}_{\rm Hard}}{d s'} &=& 
{\hat \sigma}^{\rm NC}_0(s') \,\, C_f\,\frac{\als}{\pi} \frac{1}{s^2}
\frac{s^2+s'^2}{s-s'}\,
\left[ \ln\left( \frac{s}{m^2_q}\right)- 1\right],
\eqa
where $s'$ is the invariant mass of final leptons.

One-loop radiative corrections contain terms proportional to logarithms of the 
quark masses, $\displaystyle \ln\left( \frac{s}{m^2_q}\right)$. 
They come from the initial state 
radiation contribution including virtual, soft and hard gluon emission.
These terms are in agreement with the prediction of the renormalisation group
approach, see i.e. Ref.~\cite{Kuraev:1985hb}.
In the case of hadron collisions these logarithms have been already taken 
into account in the parton density functions (PDF's). Therefore we have to 
apply a subtraction scheme to avoid the double counting. 
Linearization of the subtraction procedure
is done as described in the Ref.~\cite{Arbuzov:2005dd}.

In order to have the possibility to impose experimental cuts and event 
selection procedures of any kind, we use a Monte Carlo integration routine
based on the Vegas algorithm~\cite{Vegas}. In this case we perform a 4(6)-fold
numerical integration to get the hard gluon contribution to the partonic (hadronic)
cross sections. To get the one-loop QCD corrections we add also the
 contributions of 
the soft gluon emission and the virtual QCD loop. The cancellation of 
the dependence on the auxiliary parameter $\bar \omega$ in the sum is observed 
numerically.

For numerical evaluations we take the same set of input parameters as the one given 
in Ref.~\cite{Buttar:2006zd}. Below we consider the obtained distributions
of corrections $\delta$ for Drell--Yan CC and NC processes,  where
\vspace{-2mm}
\bqa
\delta&=&\frac{d \sigma^{1-loop}}{d \sigma^{Born}}-1 \nonumber.
\eqa
The total QCD corrections to the CC and NC cross-sections are
small, however the shape of distributions changes greatly.

\begin{figure}[!ht]
\[
\begin{array}{ccc}
\includegraphics[width=75mm,height=75mm]{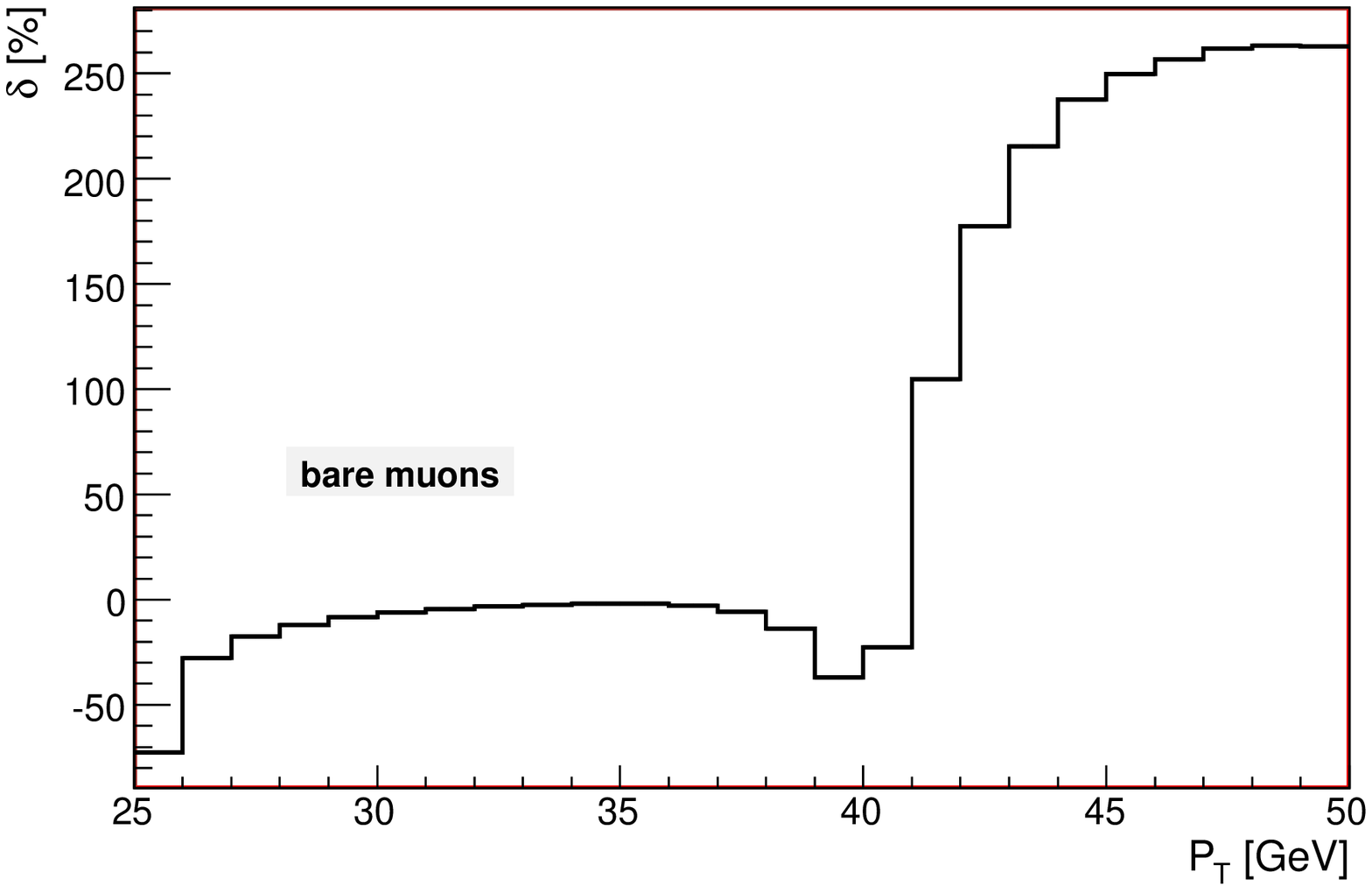} &&
\includegraphics[width=75mm,height=75mm]{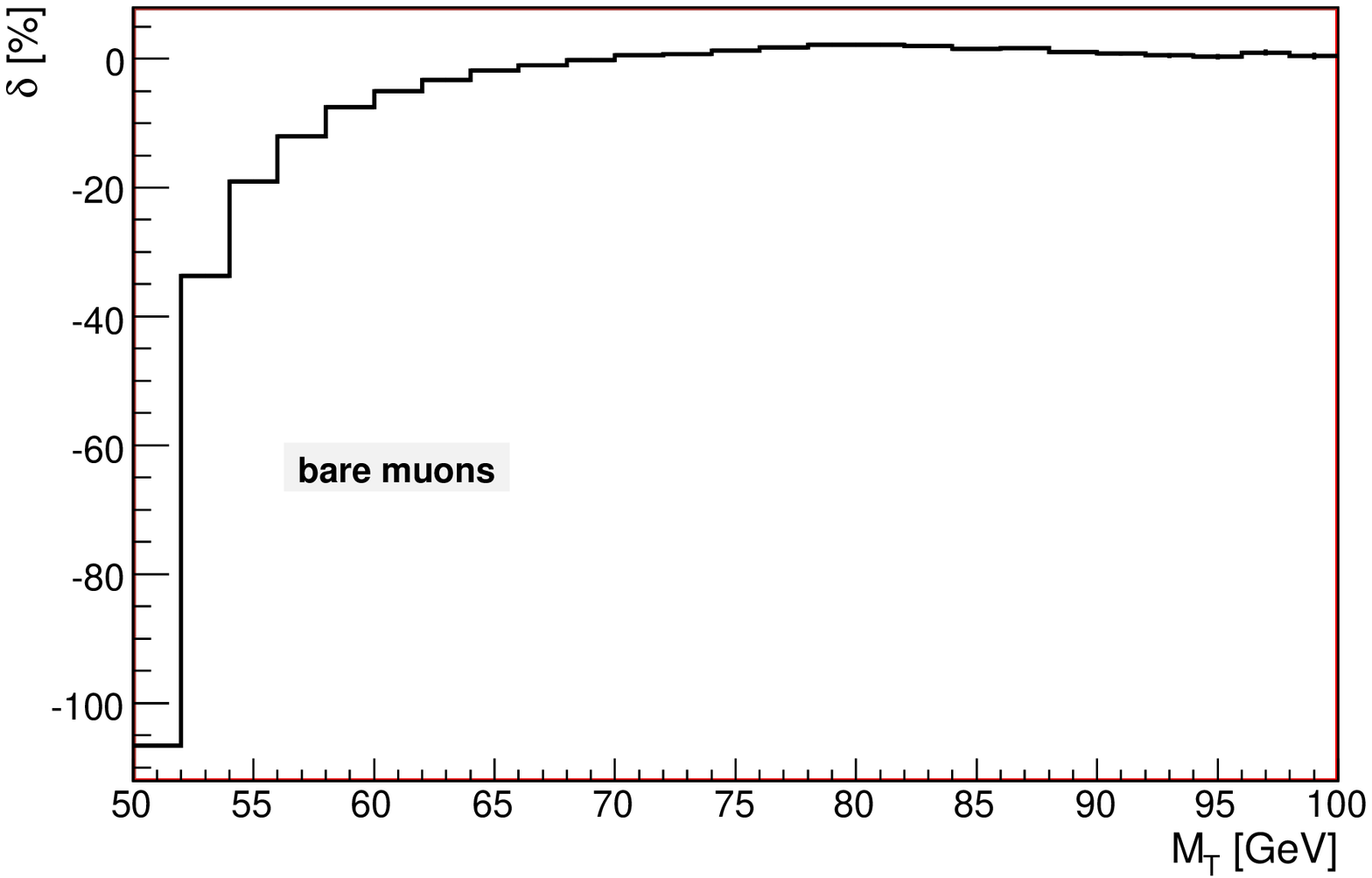} \\
a) && b)
\end{array}
\]
\vspace*{-10mm}
\caption{Transverse momentum $P_{\top}$ and transverse mass $M_{\top}$
 of neutrino--lepton pair distribution 
for Drell--Yan CC.}
\label{CCmt}
\end{figure}

Fig.~\ref{CCmt}a)  and Fig.~\ref{CCmt}b)  present the charged lepton 
transverse momentum $p^{\ell}_{\top}$  and
neutrino--lepton pair transverse mass  $M^{\ell \nu}_{\top}$ distributions
of QCD correction $\delta$ for Drell--Yan CC process,  where
$M^{\ell \nu}_{\top}= \sqrt{2 p^{\ell}_{\top} p^{\nu}_{\top} 
        \left(1-\cos \varphi_{\ell \nu}\right)}$~ 
and  $\varphi_{\ell \nu}$ is the angle in the transverse plane.
 The behaviour of distributions at left side is an edge effect -- 
near the cat we have a deficit of the hard gluon emission. Below the resonance
the correction is small. Near to the resonance $p^{\ell}_{\top}$ distribution
becomes negativ and after the resonance rapidly grows up-to 250 per sent.
But the $M^{\ell \nu}_{\top}$ distribution remains small in an wide range.

\begin{figure}[!ht]
\[
\begin{array}{ccc}
\includegraphics[width=75mm,height=75mm]{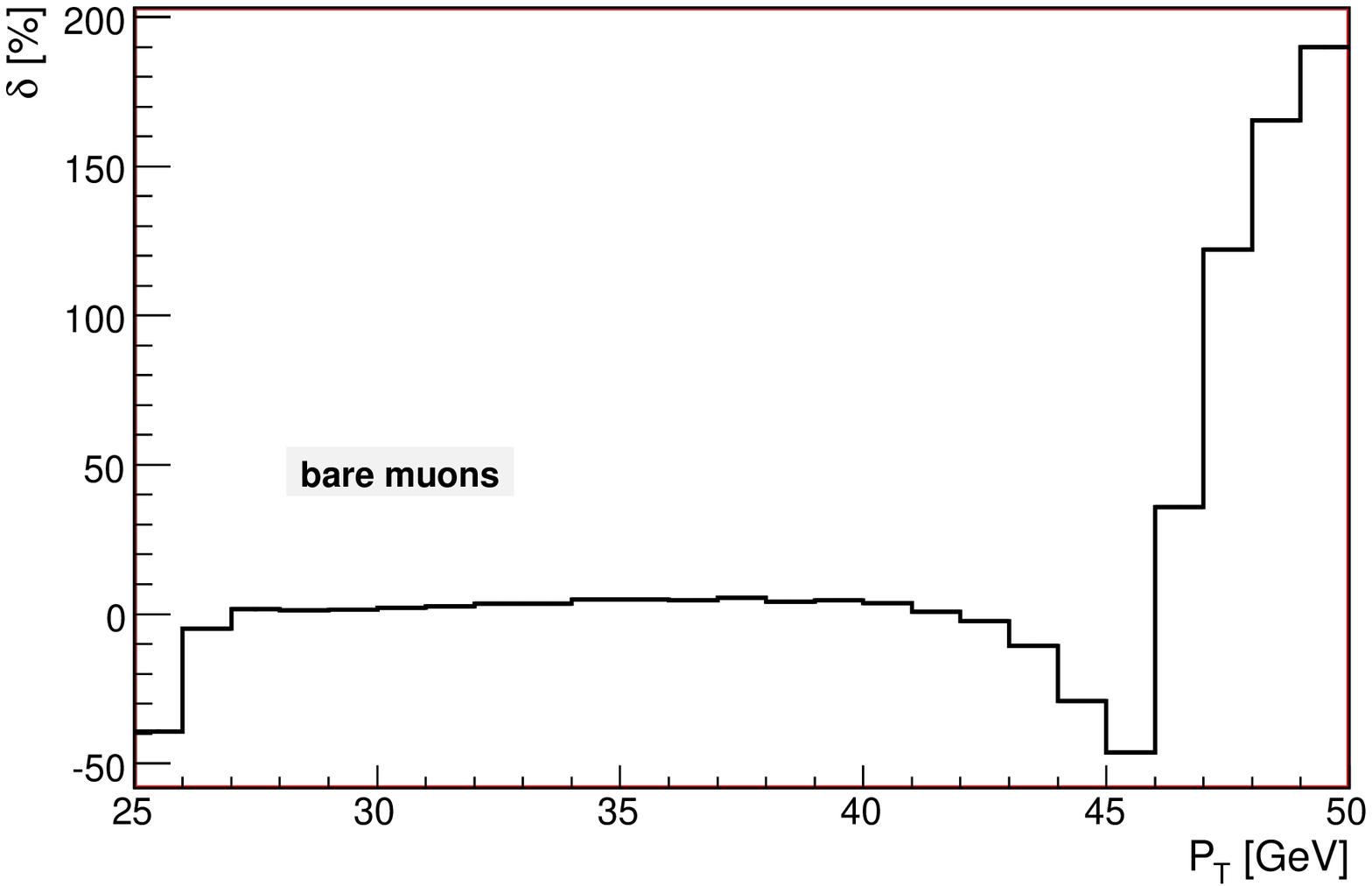}&&
\includegraphics[width=75mm,height=75mm]{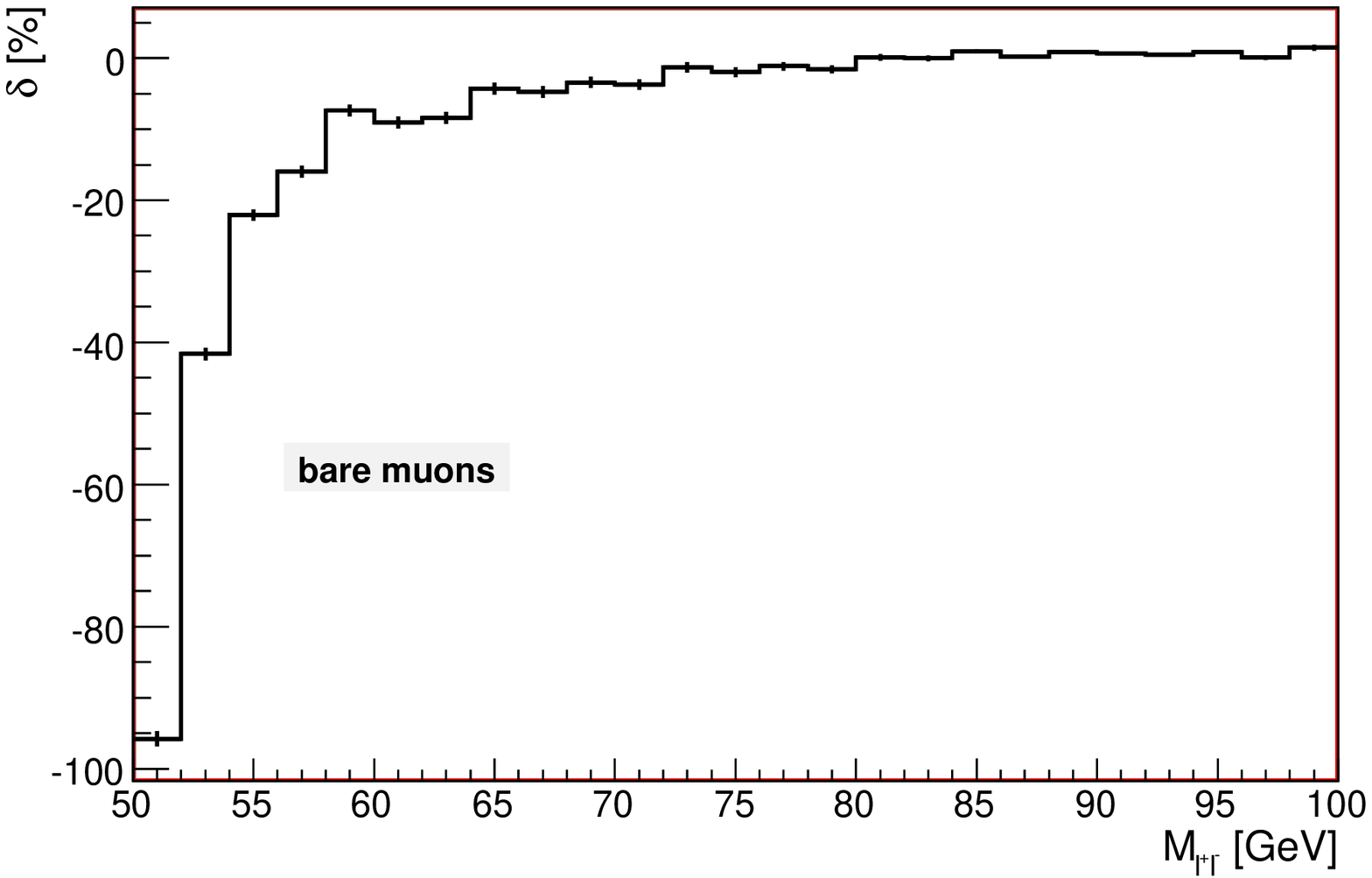}\\
a) && b)
\end{array}
\]
\vspace*{-10mm}
\caption{Transverse momentum $P_t$ and invariant mass $M_{ll}$ of lepton pair 
distributions for Drell--Yan NC.}
\label{NCfig}
\end{figure}

At  Fig.~\ref{NCfig}a) we show the lepton pair transverse momentum
distribution of QCD correction $\delta$ for Drell--Yan NC process and 
 at Fig.~\ref{NCfig}b) --
 lepton pair invariant mass  distribution. Their behaviour is similar to the 
CC case.

 Comparison of QCD and EW 
distributions was discussed in reports on workshop~\cite{CALC}.

We are going to develop a Monte Carlo event generator to describe the Drell--Yan 
processes in realistic conditions.

\section{Summary}

In this paper we described the first steps of 
creation of the QCD branch in the SANC system.
We developed environment for calculations of QCD processes. 
Then we tested successfully this environment 
by simple calculations of {\bf b2q} decays. We implemented corrections 
to top quark decays and observed that complete one--loop result differs from the 
narrow width cascade approximation result. Drell--Yan NC and CC processes were 
implemented into SANC, and we created the Monte Carlo integrators to study them.
We are going to develop unified event generator for Drell--Yan processes, 
which would include QCD and EW corrections simultaneously.
  
\section*{Acknowledgments}
We are grateful to D.~Bardin, L.~Kalinovskaya, W.~von Schlippe 
for critical reading and discussions of this paper.
This work was partially supported by the INTAS grant 03-41-1007 and by the RFBR 
grant 04-02-17192. One of us (A.A.) thanks also the grant of the President RF
(Scientific Schools 5332.2006).


\begin{thebibliography}{15}

\bibitem{SANC1.00}
A.~Andonov  et~al. {Comput. Phys. Commun.}
           {\bf 174} (2006) 481 
{{hep-ph/0411186}}.  

\bibitem{Vermaseren}
J.A.M.~Vermaseren, {\em New features of {\tt FORM},} 
{math-ph/0010025}.

\bibitem{BaVilPXX}
D.~Bardin, B.~Vilensky, P.~Christova,
Sov. J. Nucl. Phys. {\bf 53} (1991) 152.

\bibitem{BraLev} E.~Braaten, J.P.~Levelle,
{Phys. Rev.} {\bf D22} (1980) 715.

\bibitem{Bilenky} S.M.~Bilenky, {\em Introduction to Feynman diagrams and 
electroweak interactions physics}, Gif-sur-Yvette, France, Ed. Frontieres (1995).

\bibitem{Arbuzov2006}
  A.~Arbuzov, D.~Bardin, S.~Bondarenko, P.~Christova, L.~Kalinovskaya
 and R.~Sadykov, {\em SANCnews: Sector $4 f$, Charged Current}, in preparation.

\bibitem{'tHooft:1978xw}
  G.~'t Hooft and M.~J.~G.~Veltman,
  %``Scalar One Loop Integrals,''
  Nucl. Phys. B {\bf 153} (1979) 365.
  %%CITATION = NUPHA,B153,365;%%

\bibitem{Fischer2001}
M.~Fischer, S.~Groote, J.G.~Koerner and M.C.~Mauser,
{Phys. Rev.} {\bf D65} (2002) 054036.
% {{hep-ph/0101322}}.
% p.26

\bibitem{PhysPaNuc}
A.~Andonov et~al., 
Phys. Part. Nucl. {\bf 34}  (2003) 577 
[{Fiz. Elem. Chast. Atom. Yadra} {\bf 34} (2003) 1125],
{{hep-ph/0207156}}.

\bibitem{Kuraev:1985hb}
  E.A.~Kuraev and V.S.~Fadin,
  %``On Radiative Corrections To E+ E- Single Photon Annihilation At
  %High-Energy,''
  Sov. J. Nucl. Phys.  {\bf 41} (1985) 466.
  %[Yad.\ Fiz.\  {\bf 41}, 733 (1985)].
  %%CITATION = SJNCA,41,466;%%

\bibitem{Arbuzov:2005dd}
A.~Arbuzov, D.~Bardin, S.~Bondarenko, P.~Christova, L.~Kalinovskaya, G.~Nanava and R.~Sadykov,
   {Eur. Phys. J.} {\bf C 46} (2006) 407.
%  {{hep-ph/0506110}}.

\bibitem{Vegas}
 G.P.~Lepage,  {J. Comput. Phys.} {\bf 27} (1978) 192.

\bibitem{Buttar:2006zd}
  C.~Buttar et~al.,
   {\em Les Houches physics at TeV colliders 2005, Standard Model, QCD, EW, and
  Higgs working group: Summary report,}
 {hep-ph/0604120}.

\bibitem{CALC}
V.~Kolesnikov, {\em SANC: QCD sector.} 
International school--workshop ``Calculations for modern and future colliders'',
Dubna, Russia, July 15-25, 2006.

\end{thebibliography}
\end{document}